# The Triple Helix Model and the Meta-Stabilization of Urban Technologies in Smart Cities

Loet Leydesdorff [1] & Mark Deakin [2]

**Abstract**: The Triple Helix model of university-industry-government relations can be generalized from a neo-institutional model of networks to a neo-evolutionary model of how three selection environments operate upon one another. The neo-evolutionary model enables us to appreciate both organizational integration in university-industry-government relations and differentiation among functions like the generation of intellectual capital, creation of wealth, and their attending legislation. The specification of innovation systems in terms of nations, sectors, cities, and regions can then be formulated as empirical questions: is synergy generated among functions in networks of relations? This Triple Helix model enables us to study the knowledge base of an urban economy in terms of a trade-off between locally stabilized and (potentially locked-in) trajectories versus the techno-economic and cultural development regimes which work with one more degree of freedom at the global level. The meta-stabilizing potentials of urban technologies between these two levels can be used reflexively as the intelligence of a creative reconstruction making cities smart(er).

[1] Amsterdam School of Communications Research (ASCoR), University of Amsterdam, Kloveniersburgwal 48, 1012 CX Amsterdam, The Netherlands; loet@leyc.ukdesdorff.net; http://www.leydesdorff.net.
[2] School of Engineering and Built Environment, Edinburgh Napier University, Edinburgh EH10 5DT, Scotland: m.deakin@napier.ac.uk

**Introduction**

The Triple Helix model was first formulated for the study of networks of university-industry-government relations. Beyond the neo-institutional analysis of social networks, however, the Triple Helix model can be extended to a neo-evolutionary model of the dynamics in a knowledge-based economy. The three evolutionary functions shaping the selection environments of a knowledge-based economy are: (i) organized knowledge production, (ii) economic wealth creation, and (iii) reflexive control. Because reflexivity is always involved as one of them, the functions are not given, but socially constructed as the inter-human coordination mechanisms of evolving communication systems (Luhmann 1995).

In terms of network dynamics, the functions operate as selection mechanisms and thus produce densities (which can be represented as eigenvectors). From the perspective of each density, a different meaning can be provided to the events. For example, patents can be considered as output of the science system, but as input to the economy. Their third function is to provide legal protection to new ideas. Three selection mechanisms operating upon one another can be expected to generate complex dynamics (May, 1976; May & Leonard, 1975; Sonis, 2000).

In Darwin's original evolution theory, selection was first considered "natural," that is, as given. In the paradigm of evolutionary economics (Schumpeter, 1939), different selection environments were distinguished; for example, market as against non-market environments (Nelson & Winter, 1982; Von Hippel, 1988). Comparative studies across different sectors of the economy (e.g., Nelson, 1982; Carlsson, 2002 and 2006; Carlsson & Stankiewicz, 1991) and studies of different national systems of innovation (Lundvall, 1992; Nelson, 1993) have been central to this tradition. However, the analysis of interaction effects among three selection environments cannot be pursued without an analytical model (Dolfsma & Leydesdorff, 2009).

**The Triple Helix model**

In a Triple Helix model of social coordination, selection dynamics are endogenous because actors in the three institutional spheres relate reflexively. Thus, they react to each other's selections (Etzkowitz, 2008). The dynamic of this selection process is not biologically inherited (Lewontin, 2000), but cultural, i.e., dependent on the historical development of communicative competencies by the carrying agents.

Dosi (1982) already noted that two selection environments operating upon each other may generate a trajectory in a process of mutual shaping. Technological trajectories, for example, can be shaped when interfaces between markets and R&D are operating within an institutional setting. From a third perspective, specific trajectories can be considered as local actualizations in a space of possible trajectories. Three selection environments thus can provide sufficient complexity to model the techno-economic regime of a knowledge-based economy (Nelson & Winter, 1982, at pp. 258f.).

Reflexivity in inter-human communications adds another degree of freedom to this meta-biological model: the relations between the evolutionary model of interacting dynamics and the institutional layer of university-industry-government—that is, the knowledge infrastructure—are no longer one-to-one, but can historically be reconstructed. Both the differentiation among the three spheres and their interactions in networked exchange relations can change, but are also reproduced. The interacting dynamics of the relations are anchored in differentiations among the communications.

Integration and differentiation among the subsystems are concomitant: the functionally differentiated system is able to process more complexity, while exchange relations among the subsystems make it possible to change perspectives and to develop new structures at interfaces. On the one side, one can expect a configuration to be reproduced in which the generation of intellectual capital prevails within an academic environment, with wealth creation being institutionally associated with industry, while control in the public sphere can be associated with government. On the other, network relations can be expected to reflect degrees of integration, for example, in national systems.

**The meta-stability of a knowledge-based system**

Using this Triple Helix model, it becomes possible to explain the phenomena by which a knowledge-based order can be represented by means of a variety of perspectives. Each density in the network is associated with an eigenvector which positions the observable relations differently. The densities can be reproduced over time insofar as codes of communication can be developed at this next-order level of eigenvectors. Perspectives are generated as possible recombinations among the prevailing codes of communication.

One can expect more than two contexts continuously to be relevant when *discursive knowledge* is considered as a third coordination mechanism at the level of society, in addition to—and in interaction with—economic exchange relations and political control. This additional degree of freedom in the coordination provides the distinguishing feature between a knowledge-based economy and a political-economy-based account of innovation (Leydesdorff, 2006). Both institutional arrangements and functional requirements can then be deconstructed, improved, and used as leverage for reflexive reorganization. The learning capacity at the level of functions, however, is larger than at the level of institutions (Newell & Simon, 1972; Simon, 1973).

Industries have also become important producers of new knowledge, while universities and, as we shall see, the cities they increasingly come to represent, can sometimes act as organizers of regional innovation systems (Etzkowitz *et al*., 2000). The three perspectives are interwoven in social phenomena (Gieryn, 1983; Galison & Stump, 1995). As noted, patents can function in court because they offer legal protection, but they can also be used to indicate the economic value of specific knowledge products. The interactions among the dimensions of a system, however, can be analyzed with reference to the main functions of the system using, for example, factor analysis.

Gómez *et al*. (2009), for example, illustrated the third mission logic of universities by providing the following factor matrix of a set of indicators for 65 Spanish universities:

|  | Component | | | |
|---|---|---|---|---|
|  | *1* | *2* | *3* | *4* |
| *No. PhD professors* | 0.969 | | | |
| *No. students* | 0.921 | | | |
| *No. ISI publications* | 0.874 | 0.404 | | |
| *No. PhD thesis* | 0.835 | | | |
| *University age* | 0.673 | | | **-0.422** |
| *No. students/PhD prof.* | -0.62 | | | |
| *No. citations/article* | | 0.919 | | |
| *No. ISI doc/PhD prof.* | | 0.868 | | |
| *% internat. vs. Spanish publications* | | 0.818 | | |
| *% non-cited articles* | | -0.815 | | |
| *% doc. in top journals* 10* | | 0.608 | 0.398 | |
| *Input specialization (Pratt-PhD prof)* | | | 0.858 | |
| *GDP of NUTS2 regions* | | | 0.689 | **0.354** |
| *International collaboration rate* | | 0.535 | 0.627 | **-0.415** |
| *University-industry collab. Rate* | | | | **0.808** |
| *National collaboration rate* | | | | **0.605** |
| *Output specialization (Pratt-publicat.)* | -0.31 | | 0.441 | **0.516** |

**Table 1**: Rotated Component Matrix of indicators for 65 Spanish universities. Extraction Method: Principal Component Analysis. Rotation Method**:** Varimax with Kaiser Normalization. Only loadings larger than 0.3 are shown. Source: Gómez *et al*. (2009, at p. 139).

In the context of this discussion, factor 4 represents the third mission of the universities, that is, to support economic and social development. The first and second missions, that is, teaching and research, are indicated by factors 1 (26.5 %) and 2 (23.2 %), respectively. The third factor (12.5 %) indicates a correlation between relatively rich regions (in Spain) and the internationalization of research as measured in terms of coauthorship relations. Factor loadings on factor 4 show that the internationalization of research is negatively correlated with university-industry collaborations in the Spanish context. However, university-industry relations are positively correlated with national collaborations. In this dimension, university-industry relations also correlate positively with regional development and specialization (*Ibid*., p. 342).

Leydesdorff & Sun (2009) showed that in the case of Japan, university-industry coauthorship relations have declined continuously since 1980 in terms of co-authorship relations (after normalization). However, since 1994 the Japanese system has developed a new synergy between *international* co-authorship relations and *national* university-industry-government relations. The uncertainty prevailing at the national level is reduced by this international synergy. Using the neo-evolutionary Triple Helix model of a dually layered development and in terms of both institutions and functions, it remains an empirical question where and when integration or differentiation will prevail in a given configuration. The opening of China to the world market after the demise of the Soviet Union posed a major threat to the Japanese system and then the trend towards more international co-authorship at the global level could be integrated at the level of a national, regional, or even city system (Tokyo). Whether integration or differentiation prevails may vary over time and with the systems under study.

In summary, the stabilization of a local optimum can be considered as an effect of co-evolution between selections in two dimensions operating upon each other, while the third is kept relatively stable. Given a nation state, for example, national systems of innovation could be developed by interfacing political economies with techno-scientific trajectories. Competing stabilizations can also be considered as second-order variations and can further be selected for hyper-stabilization, meta-stabilization, and globalization when a third (analytically independent) selection mechanism can be specified. Hayami & Ruttan (1970) already noted this second-order selection mechanism operating on localizable stabilizations (Nelson & Winter, 1982, at p. 258). A further selection upon stabilizations can lead to globalization.

While a trajectory forms a historical trail along trade-offs, an additional (third) feedback from the environment may first induce meta-stabilization or alternatively a hyper-stabilized lock-in. Meta-stability can be considered as a condition for participation in the globalizing dimension of innovation systems because it allows universities, industry, and governments to move from the local to the global dimension, and *vice versa*.

**The articulation of three (or more) perspectives**

How can the above systems-dynamic considerations help us to understand the observable relations between the major players in a field of study? From an evolutionary perspective, the networks provide us only with instantiations of the systems (Giddens, 1984) or, more abstractly formulated, representations of the systems dynamics. The functions in the systems under study remain latent and their operations virtual when measured in terms of instances. In other words, relevant selection mechanisms which provide meaning to the events can be formulated only as hypotheses, whereas the variation in the events can be observed. The formulation of hypotheses relates measurement to theorizing. Among other things, theoretical articulation may enable us to designate latent or emerging dimensions of the systems under study.

The Triple Helix model was originally formulated as an alternative to two competing theories (Etzkowitz & Leydesdorff, 2000): one about national systems of innovation (Freeman, 1987, 1988; Lundvall, 1988, 1992; Nelson, 1993) and the second celebrating the "new production of knowledge" or "Mode-2" (Gibbons *et al*., 1994;

Nowotny *et al*., 2001). The proponents of the "Mode-2" thesis argued that the social system had undergone a radical transition that had changed the mode of knowledge production. Advocates of the "Mode-2" thesis argued that disciplinary-based knowledge would increasingly become obsolete and should be replaced with techno-scientific knowledge generated in "trans-disciplinary" projects.

Whereas this "Mode-2" model focused exclusively on transformations, the concept of national *systems* of innovations, as it prevailed in evolutionary economics, stressed the resilience of existing arrangements. Extensive research carried out in this tradition entailed systematic comparisons of different innovation systems (Nelson, 1982, 1993; Lundvall, 1992; Carlsson & Stankiewicz, 1991; Braczyk *et al*., 1998). In addition to the idea that the nation-state—as a specific construct of the 19$^{th}$ and 20$^{th}$ centuries—would provide a stable context for the development of *national* innovation systems, other scholars have sought to focus on the emergence of sectorial or regional systems as potential candidates for the stabilization of interactions among selection environments (Carlsson, 2006).

The Triple Helix model explains these differences among innovation systems in both scale and scope in terms of *possible* arrangements. Two of the three dynamics can stabilize along a trajectory when a third context remains relatively constant. Which of the three subdynamics provides a foothold may vary among instantiations and over time. When a technology is leading the trajectory along a stable path, a sectorial system can be expected to emerge (Pavitt, 1984). When governments are able to provide strong regulatory frameworks (as in the People's Republic of China) one can expect the dominance of a national system of innovation. At the regional level, trade-offs between regional governments, local universities, and industrial capacities may shape specific niches. In a niche, one may be able to construct an advantage (Cooke & Leydesdorff, 2006; Schot & Geels, 2007). However, one can expect that each niche remains in transition: a region that was able to ride a wave may be in disarray a decade later because, for example, multi-national corporations are able to buy themselves into the innovative trajectories that were stabilized at the level of the region (Beccatini *et al.*, 2003). Dynamics of scale and scope may lead to globalization, but this next-order dynamics may develop unnoticed from a local perspective.

For example, when the nations of Eastern Europe became transition economies after the demise of the Soviet Union in 1991, the ambitions of these countries to develop national systems of innovation met with interference from market forces, on the one hand, and from the ongoing political process of Europeanization, on the other. An interesting example is provided by the case of Hungary (Inzelt, 2004). Not one, but three innovation systems emerged during the transition.

A metropolitan center developed around Budapest to compete with Vienna, Munich, Prague, etc., as a seat for knowledge-intensive services, multinational corporations, etc. In the western part of the country, specific Western-European companies moved in to the extent that they were able to influence research agendas at universities. The German car manufacturer Audi, for example, developed its own university institute at a local university in a town and region in North-Western Hungary where it developed an automotive cluster (Lengyel *et al.*, 2006). A third type of innovation system was indicated in the eastern parts of the country, where traditional universities support the development of local infrastructures remaining more continuous with the old system (Lengyel & Leydesdorff, 2007).

In other words, when Hungary arrived on the European scene, it was too late to develop a purely *national* innovation system because the envisaged system was already implicated in the formation of the European Union. Transition countries became at the same time accession countries for the European Union and the resulting dynamics could henceforth only be coordinated loosely at the national level. The period for adaptation was too short for stabilizing a national system of innovations.

This "disorganization" may vary from country to country and from region to region within countries. In this case of Eastern Europe, the transition was not only a transition at the trajectory level, but a change at the regime level. Note that the nonlinear dynamics among the interacting selection environments are controlled at the level of the emerging system. This concept of "an emerging system," however, should not be reified: the interacting *uncertainties* in the distributions determine the dynamics at the systems level. One can no longer expect a stable center where decision-making can be monopolized because the one-to-one correspondence between functions and institutions no longer prevails. The fragile order of knowledge-based expectations can be updated as new knowledge becomes available. The knowledge base of a system remains a networked order of codified expectations.

This version of the Mode-2 thesis—that is, the disorganization and fragmentation of previously existing system delineations—is appreciated in the Triple Helix model in terms of a reflexive "overlay" of relations among the carriers of innovation systems (Etzkowitz & Leydesdorff, 2000). The overlay feeds back as a restructuring subdynamic on the underlying networks, and generates and/or blocks opportunities for niche-formation in a distributed mode. New competencies may be needed for further developments; new specialties are shaped as recombinations of existing disciplinary capacities. The knowledge-based dynamics are institutionally conditioned, but evolutionary in character: the reflection at the level of the overlay operates from the perspective of hindsight and can therefore be future oriented. These dynamics generate flexibilities; not as a biological process of adaptation, but as a *social* dynamics of interactions among meanings, insights, and intentions (Freeman & Perez, 1988; Leydesdorff, 2009).

From this perspective, the flexibilization and contextualization of Mode-2 is no longer confined to the knowledge production and control system (Whitley, 2001). Mergers and acquisitions in industry are increasingly knowledge-driven. The context of the European Union has changed the status of regions, and nation states can be dissolved as in the case of Czechoslovakia, or continuously reformed as in the case of Belgium. In the new regime, the system remains in "endless transition." However, this endless transition does not mean that "anything goes," but rather a continuous recombination of strengths and competitive advantages under selection pressure (Cooke & Leydesdorff, 2006). The selection processes involved are knowledge-intensive because they can only be improved by appreciating the information which comes historically available when they operate.

**The Triple Helix of Urban Technology**

From the neo-evolutionary perspective of the Triple Helix, the urban technologies of cities can be modelled as densities in networks among the three relevant dynamics of organized knowledge production, the economics of wealth creation, and governance of civil society.[3] The effects of these interactions can be expected to generate spaces—such as "structural holes" (Burt, 1995)—where knowledge can be produced and exploited to create added value. The densities of relations among the three institutional spheres in turn allow the technologies of cities to function as key components in the organization of innovation systems.

The dynamics at play in the overlay can be facilitated by the pervasive technologies of information-based communications (ICTs) currently being exploited to generate the notion of "creative cities" (Landry, 2008) and as the knowledge base of "intelligent cities" (Komninos, 2008). These "smart" technologies of cities are now being asked to work even "smart-er" (Holland, 2008). "Smarter" not just in the way they make it possible for cities to be intelligent in generating capital and creating wealth, but in entertaining models of how selection environments co-produce knowledge in innovation systems that can co-evolve with their development in possible feedback loops. How can a city participate as a node in such a network, and for what reason in which dimensions?

---

[3] This argument is drawn form a review of how the Triple Helix is being used in Europe, Latin America, and Asia (Leydesdorff and Etzkowitz, 1998). Etzkowitz and De Mello (2004), Etzkowitz *et al*. (2005), and Leta *et al*. (2006) report on the current uses of the Triple Helix model in Latin America. Leydesdorff and Sun (2009) provide examples of how the analytical framework is being used in Asia. In Europe, the model is used to generate an understanding of the knowledge-based economy, whereas attention in Latin America focuses on the deployment of the knowledge economy as a generator of democratic governance by civil society. However, the references which these accounts of the Triple Helix make to the urban economy and civil society of city-regions are limited and sometimes perfunctory.

Such a co-evolutionary mechanism for the meta-stabilization of existing institutional arrangements marks a development that takes us beyond the dismantling of national systems and construction of regional advantages, i.e., that which fall under the remit of "innovations systems" and "Mode-2" accounts. The reinvention of cities currently taking place under the so-called "urban renaissance" cannot be defined as a top-level "transdisciplinary" issue without a considerable amount of cultural reconstruction at the bottom. While clearly recognized as an important issue by advocates of the Mode-2 perspective, the highly distributed character of this reconstruction has not yet been given the consideration it demands. For accounts of this cultural reconstruction tend to reify the global perspective and fail to appreciate the meta-stable dynamics of such communications as innovations systematically worked out as the informational content of social processes operating at the local level.

In our opinion, it is the potential of this dynamic to work as such a meta-stabilizing mechanism and reflexive layer of the urban renaissance that lies behind the surge of academic interest which is currently being directed at communities as the "practical" instantiations of intellectual capital and exploitation of the knowledge produced from their organization by industrial sectors. The Triple Helix, however, adds the distinction among the codes of communication operating within these "communities of practice" and the specification of translation mechanisms among the communications (Nooteboom, 2008). We suggest the differentiation between such communications generates intellectual capital and provides new sources of the meta-stabilizing dynamic. An innovation system can use its knowledge base to counter stagnation since knowledge networks provide another—that is, analytically orthogonal, or third—selection mechanism, operating between market forces and policies.

From this perspective, national systems can also be considered as offering the opportunity for the urban renaissance to be played out on a *global* stage and for the innovation systems of such trans-national city-regions to begin reflecting the status of cities as "world class." For example, Montreal is recognized as a city particularly successful in reinventing itself and developing a "creative" force within the region (Florida, 2004; Slolarick and Florida, 2006). While informal communities are found to generate new knowledge, the city has sought to institutionalize this process of knowledge production by developing into a learning organization. This organization has managed to invent a pedagogy by which to integrate the knowledge of knowledge-intensive firms. Furthermore, this pedagogy has in turn developed the means to integrate them as key components of (e.g., regional) innovation systems.

As Cohendet and Simon (2008) have noted, it is not just universities, industry or governments, but communities that provide the environments by which it becomes possible for cities to successfully exploit the opportunity to manage such integration. Exploit it up to the point where the City of Montreal has learnt how to become a leading exponent of cultural events and known for the advantage such an innovation system manages to construct (Nowotny, 2008). In this case, the flow of cultural events into and out of intellectual capital, wealth creation, and the government of civil society interact to open up new horizons.

The only thing offered to explain the growth of Montreal as a leading exponent of cultural events from the institutional perspective has been a list of enabling conditions, such as: a strong research, development and technological community, whose shared enterprises are underpinned by leading university involvement; university involvement supported by strong leadership from the city; and a set of policies capable of governing such ventures as part of an urban regeneration program. From our neo-evolutionary perspective, however, these (and other) conditions can be hypothesized as relevant selection environments. The codes operating in these selection environments can be reproduced, adjusted, and strengthened by interacting in local settings.

The reduction of these interaction effects among relevant functions to one of the dimensions—contextualizing the other dimensions as mere conditions—tends to lead the discourse towards an overly economic representation of "innovation systems," or a singularly one-sided account of their scientific and technical qualities from the "transdisciplinary" perspective of Mode-2 knowledge production (Hessels & Van Lente, 2008). The critical distinction between the Mode-2 type accounts of creative communities, we suggest, set out by the likes of Florida (2004) and Slolarick and Florida (2006), and those of the Triple Helix model, lies in the tendency for:

- the former to remain managerial and become locked into neo-liberal policies displaying a strong entrepreneurial legacy, and then to be articulated with reference to the market economy and its regime of accumulation;
- the latter to provide a framework for analysis capable of elaborating on what the intellectual capital of universities, wealth creation of industry, and Knowledge-intensive polices have to be articulated before they can be exploited through the scientific management of the corporate strategies governing civil society's experience of such developments. Any entrepreneurial drive to by-pass the articulation of these knowledge-intensive polices fails to represent the intellectual capital, wealth creation, and governance invested in cultural constructs.

Using the Triple-Helix model, it can be recognized that cultural development, however liberal and potentially free, is not a spontaneous product of market economies, but a product of the policies, academic leadership, and corporate strategies which need to be carefully constructed as part of an urban regeneration program. Otherwise, cultural development of this kind remains merely a series of symbolic events, left without the analytical frameworks needed to explain itself in terms of anything but the requirements of the market. Any such appeal to the efficiency of the market as a means to explain cultural development can only be considered as much an analytical shortcut, holding back any meaningful specification of the policies, leadership qualities, and corporate strategies underpinning an urban regeneration program.

Cities like Montreal and Edinburgh show how the creative ecology of an entrepreneurship-based and market dependent representation of knowledge-intensive firms can be replaced with a community of policy makers, academic leaders, and corporate strategists. Alliances that in turn have the potential to liberate cities from the stagnation which they have previously been locked into and offer communities the freedom to develop polices, with the leadership and strategies, capable of reaching beyond the idea of "creative slack" as a residual factor. For in order to be more than intelligent and smart, and in that sense, "smarter," cities need the intellectual capital required to not only meet the efficiency requirements of wealth creation under a market economy, but to become *centres of creative slack* distinguished by virtue of their communities having the political leadership and strategies which are capable of not only being culturally creative, but enterprising in opening-up, reflexively absorbing, and discursively shaping, both the economic and governmental dimensions of corporate management.

The neo-evolutionary analysis guides us towards the intellectual capital of such creativity by focusing attention on those dimensions of corporate management making it possible for urban regeneration programs to function as meta-stabilizing mechanisms underpinning civil society's integration of cities into emerging innovation systems (Deakin and Allwinkle, 2007; Deakin, 2008, 2009a). The significance of this knowledge-based reconstruction in turn resting in the real capacity a meta-stabilization that remains not only cultural, but political and economic insofar as such mechanisms enable urban regeneration programs to function as systems of innovation responding to the "creative destruction" of the global and "reflexive reconstruction" at the local level. The "creative reflexivity" of this meta-stabilisation is far from "symbolic", or of merely representational significance insofar as it generates the critical reinforcement needed to communicate the democratic values required for civil society to govern over any such "programmatic" integration of cities into emerging innovation systems (Deakin, 2009a and b). Seemingly elusive concepts such as innovation systems then can be entertained as hypothetical, yet cultivated as informed alternatives to already existing frames of reference.

Without cognitive deconstruction and analysis, cultural events run the risk of being reified to little more than signifiers of a market economy. However, a reflexive turn allows the "best practice" examples to be evaluated—instead of imitated—in terms of functional advantages. While there is no single "best practice" from an evolutionary perspective, this is not critical as long as there is sufficient "slack" in the environment to learn from failures and accommodate alternatives in ways which offer the prospect of self-regenerating actions at the level of the network. The different functionalities this produces can then in turn be articulated into specific policies, informed and further improved by learning from what works and uncovering the reasons why.

Such a critical approach—as compared to the boosting of self-proclaimed "best practices"—challenges policy makers to raise additional questions such as: whether the university should participate in such an integration, and if so how? What other potentials exist, but have hitherto been insufficiently articulated towards industry? For example: should technologically oriented faculties only be involved, or might this also include the social sciences? These questions arise because here the technology of city-regions surfaces for what it can rightly be considered to be: "nested centres" of control, dependent for their further economic and social development not only on the market, but intellectual capital and wealth creation capabilities of reflexive and self-organizing systems.

**The normative implications of the model**

While the neo-institutional model of intersecting networks may guide the researcher towards instances where university-industry-government relations can be studied empirically, in the neo-evolutionary model the emphasis remains on finding explanations for the dynamics of knowledge-based systems. An *absence* of relations (e.g., in structural holes) can be just as important as their presence given the specification of an expectation. In this design the focus shifts methodologically to empirical distributions or, in other words, testing observable variation against the theoretical specification of selection mechanisms.

Because a one-to-one relation between institutional agency and functions in a network can no longer be assumed, the relevant contexts of institutional agents need to be specified as functional requirements. The specification of functions provides yardsticks for the measurement from a systems perspective. For example, one can raise the question in which respects a Technology Transfer Office also filters information at the interface which it is intended to stimulate. From an institutional perspective, it is more difficult to raise such a research question because institutional interests are also involved.

Although the model's primary purpose is to help specify a research agenda, the Triple Helix thesis has also been used for neo-corporatist and neo-liberal agendas of policy making. The Swedish state agency for innovation, Vinnova, for example, has made "The Triple Helix" its official strategy (Etzkowitz, 2008) in accordance with this country's neo-corporatist traditions. According to others (e.g., Mirowski & Sent, 2007),[4] a further commercialization of the university could result from this "ideology." However, while the institutional analysis serves to place the "opening-up of the black box" of systematic knowledge production on an agenda, this dynamics is not yet analyzed further in terms of its specific effect on and potentials for the resulting innovation systems (Rosenberg, 1982; Whitley, 1984).

For example, in the "Varieties of Capitalism" debate Hall & Soskice (2001) neglected the knowledge production function of civil society as an independent source of variance and focused almost exclusively on differences in political economies. Similarly but *mutatis mutandis*, "best practices" in university-industry relations cannot be studied for their transferability among regions for the simple reason the regulatory and legislative conditions underlying the role of government is subject to different legal and cultural criteria.

Three, instead of two, analytically different selection mechanisms are involved in knowledge-based systems. The Triple Helix model stimulates the researcher to discuss all three functions in a research design and thus to enrich the explanation. For example, Van Looy *et al*. (2007) showed that the introduction of Bayh-Dole type legislation had an independent effect on patenting by universities when compared among European nations. The increases were shown to range from 250% for Germany, or 300% for Belgium, to 500% for Denmark. More detailed analysis in the Belgian case revealed that the university has to ensure that inventive activities do not jeopardize research and education. In addition, each university has to install

---

[4] Mirowski & Sent (2007) replace the Triple Helix categories with "Corporate," "Governmental," and "Educational" (CGE). In their opinion, CGE fits better in the "hegemonial discourse" of science and technology studies, while the Triple Helix employs a terminology used by "scholars from the periphery."

procedures to ensure a fair return on investment in patenting for researchers and research groups.

In other words, the neo-evolutionary version of the Triple Helix model does not prescribe that one "should" institutionally collaborate in local networks and cities "ought" to develop programs in the service of regional innovation systems. What the model suggests is that a three-dimensional design is sufficiently complex to analyze the integration and differentiation mechanisms which exist among the sub-dynamics of a knowledge-based system. One may wish to add more dimensions of analysis (as in the Leydesdorff & Sun's (2009) study of Japan). The analysis of a complex system in terms of a single "co-evolution" or "mutual shaping" between two dynamics, however, tends to underestimate the complexity of the regimes in knowledge-based systems by focusing on historical trajectories of their integrations. A co-evolution, for example, may bifurcate and reproduce functional differentiation at a later stage (Dolfsma & Leydesdorff, 2009; Geels *et al*., 2008).

In order to reach beyond its (e.g., geographical) borders, the urban technology of city-regions requires the unfolding of such a rich set of discursive reflections and reflexive analysis of their "regenerative effects" on the intellectual capital of wealth creation under the governance of civil society. Having said this, there remains an ever present danger of under-representing relevant discursive domains because of the pressure for change from "outside" agencies (Deakin, 2008, 2009a and b). Differentiation of functions reduces such pressure because differences in positions and missions can also be appreciated from the inside and as components of evolving innovation systems. The knowledge-based economy is based just as much on how an innovation system performs as the producer of discursive knowledge as anything else and, therefore, needs the reflexivity and self-organizing tendencies of a Triple-Helix model. The model suggests that the unfolding of such dimensions and perspectives offers the reflexivity and self-organizing properties which enable agents to "turn innovation inside-out" and "manage" this by participating in actions purposefully designed (as programs) to help shape the form, content, and directions their development may take.

**Conclusions**

This paper has set out to demonstrate how the Triple Helix model enables us to study the knowledge base of an urban economy in terms of civil society's support for the evolution of cities as key components of innovation systems. Cities can be considered as densities in networks among three relevant dynamics in the intellectual capital of universities, industry of wealth creation and their participation in the democratic government of civil society. The effects of these interactions can generate spaces and dynamics within cities where knowledge exploration can also be exploited. The densities of relations among the spaces of the three institutional spheres enable cities to bootstrap the technology of regional innovation systems.

These technologies, we have argued, are enabled by the all-pervasive technologies of information-based communications (ICTs) currently being exploited to generate the notion of "creative cities," as the knowledge base of intelligent cities and their augmentation into smart(er) cities. "Smart(er)" at exploiting information and communication technologies that are not only creative, or intelligent in generating intellectual capital and creating wealth, but smart-er in the sense which the selection environments governing their knowledge production make it possible for cities to become integral parts of emerging (e.g., regional) innovation systems. The specificity of possible matches is not given, but remains constructed, reflexively accessible, knowledge-intensive, and fragile due to the fact that discursive knowledge remains based on representations which can be further informed.

This reflexive instability of a knowledge-based system provides the co-evolutionary mechanism between institutional stabilization and communicative meta-stabilization which offers us the possibility of relating the city to next-order systems in a process of globalization. The capacity to process this transition reflexively, that is, in terms of translations, marks a development which takes us beyond the dismantling of national systems and construction of regional advantages. Using this Triple-Helix model, it can be appreciated that cultural development, however liberal and potentially free, is not a spontaneous product of market economies, but the outcome of a set of policies, academic leadership qualities, and corporate strategies, which all need to be carefully reconstructed, pieced together, and articulated before management can govern over them as requirements of an urban regeneration program.

Gómez, I., M. Bordons, M. T. Fernández, and F. Morillo. 2009. Structure and research performance of Spanish universities. *Scientometrics,* 79(1): 131-146.

Hall, P. A., and D. W. Soskice, eds. 2001. *Varieties of Capitalism: The Institutional Foundations of Comparative Advantage*. Oxford, etc.: Oxford University Press.

Hayami, Y., and V. W. Ruttan. 1970. Agricultural Productivity Differences among Countries. *The American Economic Review,* 60(5): 895-911.

Hessels, L., & H. van Lente. 2008. Re-thinking new knowledge production: A literature review and a research agenda. Research Policy, 37(4): 740-760.

Inzelt, A. 2004. The evolution of university–industry–government relationships during transition. *Research Policy,* 33(6-7): 975-995.

Komninos, N. 2008. *Intelligent Cities and Globalisation of Innovation Networks* London, Taylor & Francis.

Landry, C. 2008. *The Creative City*, London, Earthscan.

Lengyel B., E. Lukács, and G. Solymári. 2006. A külföldi érdekeltségű vállalkozások és az egyetemek kapcsolata Győrött, Miskolcon és Szegeden. *Tér és Társadalom,* 4: 127-140.

Lengyel, B., and L. Leydesdorff. 2007. *Measuring the knowledge base in Hungary: Triple Helix dynamics in a transition economy.* Paper presented at the 6th Triple Helix Conference, 16-19 May 2007, Singapore.

Leydesdorff, L. 2006. *The Knowledge-Based Economy: Modeled, Measured, Simulated*. Boca Raton, FL: Universal Publishers.

Leydesdorff, L. 2009. The Non-linear Dynamics of Meaning-Processing in Social Systems. *Social Science Information,* 48(1), 5-33.

Leydesdorff, L., and H. Etzkowitz. 1998. The Triple Helix as a model for innovation studies. *Science and Public Policy,* 25(3): 195-203.
28